%% file: PaperAnisPMB_Nov2018.tex
\documentclass[1p]{elsarticle}

\usepackage{lineno,hyperref}

\expandafter\let\csname equation*\endcsname\relax
\expandafter\let\csname endequation*\endcsname\relax
\usepackage{amsmath}
\usepackage{amsfonts}
\usepackage{ulem}
\usepackage{empheq}
\usepackage{pgfplots}
\usepackage{graphicx}
\usepackage{subfigure}

\usepackage{color}

\definecolor{pinegreen}{rgb}{0.0, 0.47, 0.44}

\journal{Physics in Medicine and Biology}

\biboptions{sort & compress}

\begin{document}

\begin{frontmatter}

\title{Handling Anisotropic Conductivities in the EEG Forward Problem with a Symmetric Formulation}

\author[Brest]{Axelle Pillain}
\author[Brest]{ Lyes Rahmouni } 
\author[Torino,Brest]{Francesco Andriulli\corref{mycorrespondingauthor}}
\address[Brest]{Computational Electromagnetics Research Laboratory, IMT Atlantique, Brest, France}
\address[Torino]{Politecnico di Torino, Turin, Italy}


\cortext[mycorrespondingauthor]{Corresponding author}
\ead{francesco.andriulli@imt-atlantique.fr}


\begin{abstract}
The electroencephalography (EEG) forward problem,  the computation of the electric potential generated by a known  electric current source configuration in the brain, is a key step of 
\textcolor{black}{EEG source analysis}. In this problem, it is often 
\textcolor{black}{desired} to model the anisotropic conductivity profiles of the skull and of the white matter. These profiles, however, cannot be handled by standard surface integral formulations and the use of volume finite elements is required.
\textcolor{black}{. L}everaging on the representation theorem using an anisotropic fundamental solution, \textcolor{black}{this paper} proposes a modified symmetric formulation for solving the EEG forward problem by a surface integral equation which can take into account anisotropic conductivity profiles.
A set of numerical results is presented to corroborate theoretical treatments and to show the impact of the proposed approach on both canonical and real case scenarios.
\end{abstract}


\end{frontmatter}


\section{Introduction}

Epileptic source localization from high resolution electroencephalographies (EEGs) is a fundamental step in the pre-surgical evaluation of focal epileptic patients that are refractory to pharmacological treatments and for whom surgical resection of the epileptic focus is considered \cite{lantz2003epileptic, plummer2008eeg,michel2012towards,grouiller2016presurgical}.
In this brain imaging technique, starting from the electrical potential measured on the scalp, the brain current sources responsible for focal epilepsy are localized by solving an inverse source problem  \cite{grech2008review,becker2014eeg, gramfort2012mixed}.
Solving this inverse problem requires \textcolor{black}{the} multiple solution\textcolor{black}{s} of an EEG forward problem that provides, from known brain electrical current sources, the surface potential measured at the electrodes' locations \cite{grech2008review,hallez2007review}.
The solution of the forward problem must be computed with the highest possible precision \cite{fuchs2001boundary,oostenveld2002validating,acar2013effects} to increase the accuracy of the source localisation process.

In solving the EEG forward problem, spherical head models have been historically used since analytic solutions are available for them  \cite{de1988potential,zhang1995fast}. However, modern techniques rely on the use of realistic head models that require a numerical solution, but for which the accuracy of the forward EEG solution is largely improved \textcolor{black}{\cite{cuffin1996eeg, haueisen1997influence, baillet2001evaluation,ramon2006influence,vorwerk2014guideline}. Several methods can be used for numerically solving the EEG forward problem, including Finite Difference (FDM) \cite {vanrumste2001validation, nixon2003numerical} , Finite Element (FEM) \cite {wolters2004efficient} and Boundary Element methods (BEM) \cite{Hamalainen-1989,Fuchs-1998,kybic2005common,stenroos2012bioelectromagnetic}. Boundary Element methods have been quite popular given that they require only surfacic discretization of the brain layers when compared to the other two techniques that rely on volume discretizations. These methods however are not panacea given that, in their standard incarnations, they cannot handle anisotropic conductivity profiles. Indeed, correct modeling of anisotropic conductivity profiles is quite important given the influence of white matter and skull anisotropic conductivities on source localization \cite{marin1998influence,wolters2006influence,gullmar2010influence,haueisen2002influence,
rullmann2009eeg, aydin2014combining, montes2014influence}. The reader should notice that anisotropic conductivity profiles can be naturally treated with FEM approaches. 
Even if the computational load can be reduced by resorting to transfer \textcolor{black}{matrices}, as in \cite{weinstein2000lead} or \cite{genccer2004sensitivity}, a purely surfacic BEM method (with anisotropic modeling capabilities) could be desirable. Indeed, in solving the EEG forward problem, the integral methods, which are based on an operator having an asymptotically smooth kernel, can easily be used with a fast multipole method (FMM) \cite{kybic2005fast}\textcolor{black}{. This decreases} their complexity to a quasi-linear one ($O(N \log(N))$, where $N$ is the number of unknowns). Moreover, exploiting the reciprocity theorem as in \cite{vanrumste2001validation, ziegler2014finite}, further reduces the computational burden when solving the EEG inverse problem. Some steps in building an anisotropic BEM formulation have been presented in \cite{olivi2011handling} where 
virtual sources are placed in the fibers to account for their anisotropic properties. Moreover, \cite{zhou1994application} proposed an interesting coordinate transform to handle constant anisotropy in a single conducting body. No details, however, were provided for the multi-compartments problem and the associated numerical solutions.} \textcolor{black}{ Preliminary and partial results on an indirect anisotropic formulation for the EEG forward problem had been presented in the conference contributions \cite{pillain2015indirect,pillain2016conf}. }

This work proposes a new \textcolor{black}{approach to the problem that falls under the family of symmetric formulations as originally introduced in \cite{kybic2005common} and results in a} surfacic discretized BEM formulation capable of handling constant piecewise homogeneous conductivity profiles including anisotropies for both nested and non-nested compartments. 
This is obtained by leveraging on the representation theorem for the Poisson equation, using an anisotropic fundamental solution. Instead of building a harmonic solution as is commonly done in building EEG integral formulations, our approach directly computes the unknown potential. This allows to enforce boundary conditions also across different anisotropic media. This approach also naturally permits to take into account non-nested domains as in \cite{kybic2006generalized}, 
\textcolor{black}{and it therefore renders possible further subdivisions of the compartments according to the variation of their conductivity tensors}.

The article is organized as follows: Section \ref{Backgrd} describes background material and sets the notations. Section \ref{BEMFormAn} presents the new anisotropic \textcolor{black}{symmetric} integral equation. Section \ref{Discr} focuses on the discretization of the BEM solution \textcolor{black}{and its implementation}. Section \ref{NumRes} complements all the theoretical developments with numerical results that show the effectiveness of the newly proposed method in both canonical and real case scenarios.

\section{Background and Notation}\label{Backgrd}

\begin{figure}
	\centering
	{\includegraphics[width=7cm]{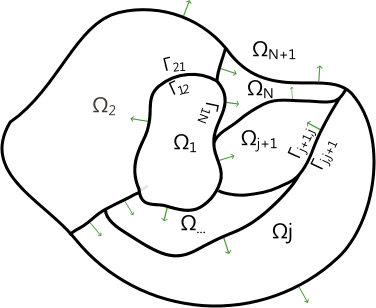}}
	\hfill
	\caption{Decomposition of the domain $\Omega$ into subdomains $\Omega_i$ with boundary $\Gamma_{ij}$ and normal $\vec{n}$.}
	\label{fig:NotnestedDom} 
\end{figure}

\textcolor{black}{Consider a domain $\Omega$ divided into $N$ non overlapping subdomains $\Omega_i$ with Lipschitz boundary $\partial_{\Omega_i}$ such that $\Omega = \bigcup_j^N \overline{\Omega}_i$. The external domain, $\mathbb{R}^3 \backslash \Omega$, is denoted with $\Omega_{N+1}$. We further define 
 $\Gamma_{ij} = \overline{\Omega}_i \cap \overline{\Omega}_j$ the interface beween $\Omega_i$ and $\Omega_j$ as well as $\omega _i$ the set of neighboring domains of $\Omega_i$ so that  $\omega_i = \{ j \leq N+1  \vert  \overline{\Omega}_i \cap \overline{\Omega}_j \neq \emptyset \}$. Then $\partial_{\Omega_i}$ can be defined as $\partial_{\Omega_i} =  \bigcup_{k\in I_i} \Gamma_{k}$ with  $I_i = \{ k \vert \Gamma_k = \Gamma_{ij}, j \in \omega_i \}$, the set of interfaces composing $\partial_{\Omega_i}$. The normal $\vec{n}_{ij}$ to the interface $\Gamma_{ij}$ is oriented from $\Omega_i$ to $\Omega_j$ with $i<j$.} Figure \ref{fig:NotnestedDom} shows a general decomposition of $\Omega$ into subdomains. Standard EEG BEM formulations often  make use of nested domains \cite{kybic2005common}, so that when a three layers geometry is chosen, the compartments represents the brain, the skull and the scalp. In this particular case, $\partial_{\Omega_i} = \Gamma_{i, i-1} \cup \Gamma_{i, i+1}$. 
We define the traces of a function $g$ on a boundary $\Gamma_i$ and of its conormal derivative as \cite{sauter2011boundary}
\begin{subequations}\label{traces}
	\begin{align}
&\gamma_{0i}^\pm g = g_{|\Gamma_i^{\pm}}\label{trace0}\\
&\gamma_{1_{i}}^\pm g = \vec{n} \cdot \bar{\bar{\sigma}} \nabla g_{|\Gamma_i^{\pm}}\label{trace1}.
\end{align}
\end{subequations}
where $\bar{\bar{\sigma}}$ is the conductivity tensor. Moreover $[\cdot]_{k}$ will denote the jump of a function across the surface $\Gamma_{k}$: $[g]_{k}  = \gamma_{0k}^- g_{k} - \gamma_{0k}^+ g_{k} $ and $\left[\vec{n} \cdot \bar{\bar{\sigma}} \nabla g\right]_{k}  = \gamma_{1k}^- g_{k} - \gamma_{1k}^+ g_{k} $. 

The EEG forward problem amounts at computing the electric potential $V$ knowing the brain electric sources $f = \nabla \cdot j$ when the current $j$ propagates in a medium of conductivity $\bar{\bar{\sigma}}$ which is a real symmetric and positive definite matrix. Under standard quasi-static assumptions this calls for the solution of the Poisson's equation \cite{sarvas1987basic}
\begin{equation}\label{fwdPb}
\nabla \cdot \bar{\bar{\sigma}} \nabla V = f
\end{equation}
with boundary conditions at each interface $\Gamma_{k}$
\begin{subequations}\label{boundCond}
	\begin{align}
& \left[V\right]_{k} = 0 \; \label{Cont} \\
& \left[\vec{n} \cdot \bar{\bar{\sigma}} \nabla V\right]_{k}=0 \;\label{ContDer}.
\end{align}
\end{subequations}
The  conditions above enforce the continuity of the potential and of its derivative between different compartments of the head.
The conductivity $\bar{\bar{\sigma}}$ is assumed to be piecewise homogeneous and potentially anisotropic, $\bar{\bar{\sigma}}_i$ will denote the conductivity of the domain $\Omega_i$.
The source term $f = \nabla \cdot j$ is usually a linear combination of dipole sources $f_j$ such that $ f_j = q_j\cdot\nabla \delta_{r_{0j}}$ with $q_j$ the dipole moment and $r_{0j}$ the dipole position \cite{deMunck1988dip},  \cite{nunez2006electric}.

\section{A Surface Integral Formulation for Anisotropic Conductivity Profiles} \label{BEMFormAn}

The fundamental solution (Green's function) of \eqref{fwdPb} in an unbounded medium reads \cite{cances1997new}
\begin{equation}\label{GreenFct}
G_j(r,r') = \frac{1}{4\pi} \frac{1}{\sqrt{det(\bar{\bar{\sigma}}_j)}\sqrt{\bar{\bar{\sigma}}_j^{-1}(r-r')\cdot(r-r')}}
\end{equation}
where $\bar{\bar{\sigma}}_j$ denotes the homogeneous conductivity tensor of the domain $\Omega_j$.
\textcolor{black}{Decomposing the potential $V$ into different contributions in each domains $\Omega_i$, $V = \sum_{i=1}^{N} V_i$ such that $V_i(r) = V(r)$ if $r\in \Omega_i$ and 0 elsewhere, the representation theorem in $\Omega_i$ gives \cite{sauter2011boundary}
\begin{equation}\label{eq:RprTh}
V_i(r) = \int_{ \Omega_i} G_i(r,r') f(r')dr' + \int_{ \partial_{\Omega_i}} \bar{\bar{\sigma}}_j \vec{n'} \cdot \nabla G_i(r,r')V(r')dr' - \int_{ \partial_{\Omega_i}} \bar{\bar{\sigma}}_j \vec{n'}\cdot \nabla V(r') G_i(r,r') dr'.
\end{equation}
Define the following integral operators
\begin{equation}\label{singleLOpOm}
\mathcal{S}_i \nu(r) = \int_{\partial_{\Omega_i}} G_i(r,r') \nu(r') d r' \quad \mathcal{S}_i : H^{-1/2}(\partial_{\Omega_i}) \rightarrow H^{1}(\Omega)
\end{equation}
and
\begin{equation}\label{doubleLOp}
\mathcal{D}_i \mu(r) = \int_{\partial_{\Omega_i}} \mu(r') \vec{n'} .(\bar{\bar{\sigma_i}} \nabla G_i(r,r') )  d r' \quad \mathcal{D}_i : H^{1/2}(\partial_{\Omega_i}) \rightarrow H^{1}(\Omega).
\end{equation}
Here and in the following $H^{s}$  will denote the Sobolev space of order $s$. Then, since $\bar{\bar{\sigma_i}}$ is symmetric, the representation theorem \eqref{eq:RprTh} reads
\begin{equation}\label{eq:RprThOp}
V_i(r) = \int_{ \Omega_i} G_i(r,r') f(r')dr' + \mathcal{D}_i V(r) - \mathcal{S}_i  \vec{n}\cdot \bar{\bar{\sigma}}_i \nabla V(r').
\end{equation}}
\textcolor{black}{Note that, with our notations and normal orientation, \cite{sauter2011boundary}\\
\begin{minipage}{.45\textwidth}
	\begin{subequations}
		\begin{empheq}[left=\empheqlbrace]{align}
			&\gamma_{0i}^{\pm}\mathcal{S}_j = \sum_{k  \in I_j} S_{ik, \Omega_j}\\
			&\gamma_{1i}^{\pm}\mathcal{S}_j = \mp \frac{1}{2} + \sum_{k \in I_j} D^*_{ij, \Omega_j}
		\end{empheq}
	\end{subequations}
\end{minipage}
\begin{minipage}{.08\textwidth}
	\centering
	and
\end{minipage}
\begin{minipage}{.45\textwidth}
	\begin{subequations}
		\begin{empheq}[left=\empheqlbrace]{align}
			&\gamma_{0i}^{\pm}\mathcal{D}_j =  \mp \frac{1}{2} + \sum_{k \in I_j} D_{ij, \Omega_j}\\
			&\gamma_{1i}^{\pm}\mathcal{D}_j = \sum_{k \in I_j} N_{ij, \Omega_j}
		\end{empheq}
	\end{subequations}
\end{minipage}\\
 with, 
\begin{subequations}\label{Op}
\begin{align}
& S_{ik, \Omega_j} \nu(r) = \int_{ \Gamma_k} G_j(r,r') \nu(r') d r', \quad S_{ij} : H^{-1/2}(\Gamma_k) \rightarrow H^{1/2}(\Gamma_i),\label{singleLOp}\\
& D^*_{ik, \Omega_j}\nu(r) = \int_{ \Gamma_k} \nu(r') \vec{n} \cdot(\bar{\bar{\sigma}}_{j} \nabla G_j(r,r') )d r', \quad D^*_{ij} : H^{-1/2}(\Gamma_k) \rightarrow H^{-1/2}(\Gamma_i),\label{D*Op}\\
& D_{ik, \Omega_j}\mu(r) = \int_{ \Gamma_k} \mu(r') \vec{n'} \cdot(\bar{\bar{\sigma}}_{j} \nabla G_j(r,r') )d r', \quad D_{ij} : H^{1/2}(\Gamma_k) \rightarrow H^{1/2}(\Gamma_i), \text{ and } \label{DOp}\\
& N_{ik, \Omega_j}\mu(r) = \vec{n} \cdot\bar{\bar{\sigma}}_{i} \nabla \int_{ \Gamma_k} \mu(r') \vec{n'} \cdot(\bar{\bar{\sigma}}_{j} \nabla G_j(r,r') )d r', \quad N_{ij} : H^{1/2}(\Gamma_k) \rightarrow H^{-1/2}(\Gamma_i)\label{NOp}.
\end{align}
\end{subequations}}

\begin{figure}\label{interfaceDescrpt}
	\hfill
	\subfigure[Internal problem in $\Omega_1$.\label{fig:label:Omega1}]{\includegraphics[width=0.32\textwidth]{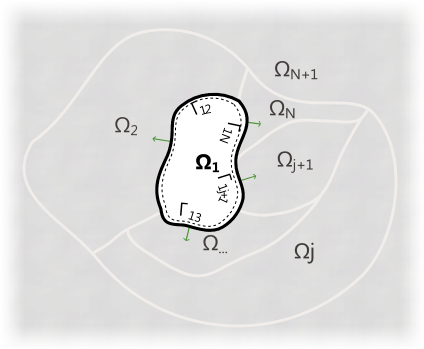}}
	\hfill
	\subfigure[Internal problem in $\Omega_2$.\label{fig:label:Omega2}]{\includegraphics[width=0.32\textwidth]{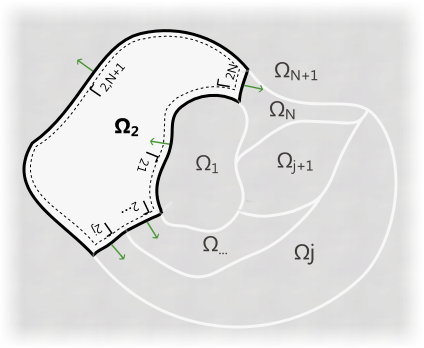}}
	\hfill
	\subfigure[Interface problem between $\Omega_1$ and $\Omega_2$.\label{fig:label:Interface12}]{\includegraphics[width=0.33\textwidth]{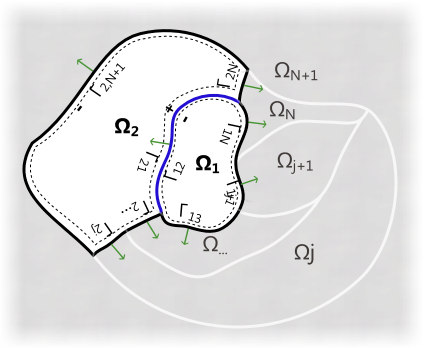}}
	\hfill
	\caption{Conventions used in setting up the integral equations. 
	}
	\label{fig:domainDec}
\end{figure}

\textcolor{black}{Now, let us consider $\Gamma_{k} = \overline{\Omega_i} \cap \overline{\Omega_j} \neq \emptyset$ with $i<j$ to express the interior problems and combine the obtained equations to enforce the boundary conditions at the interface $\Gamma_{k}$ (see  figures~\ref{fig:label:Omega1},~\ref{fig:label:Omega2}, and~\ref{fig:label:Interface12}). The first step is to express also the interior problem \eqref{eq:RprThOp} in $\Omega_j$
\begin{equation}\label{eq:RprThOpJ}
V_j(r) = \int_{ \Omega_j} G_j(r,r') f(r')dr' + \mathcal{D}_j V(r) - \mathcal{S}_j \vec{n}\cdot \bar{\bar{\sigma}}_j \nabla V(r).
\end{equation}
Then, decomposing  $V_{j_{\vert \partial\Omega_j}}$ and $V_{i_{\vert \partial\Omega_i}}$ into their contributions on the different interfaces $\Gamma_l$ with $l\in I_j$ or $I_i$ respectively, such that 
$ V_{j_{\vert \Gamma_l}} = \mathcal{V}_{l} $ for $l \in I_j$ or $ V_{i_{\vert \Gamma_l}} = \mathcal{V}_{l}$ for $l \in I_i$ respectively, and taking the trace of \eqref{eq:RprThOpJ} and of \eqref{eq:RprThOp}
we obtain
\begin{equation}\label{eq:Gamma0RprThOpJ}
\gamma_{0_{k}}^+V_j = \gamma_{0_{k}}^+v_{dip_j} + \sum_{l \in I_j} D_{kl, \Omega_j}  \mathcal{V}_{l} + \frac{1}{2} \mathcal{V}_{k} + \sum_{l \in I_j}S_{kl, \Omega_j} \vec{n}\cdot \bar{\bar{\sigma}}_j \nabla  \mathcal{V}_{l}.
\end{equation}
\begin{equation}\label{eq:Gamma0RprThOp}
\gamma_{0_{k}}^-V_i = \gamma_{0_{k}}^-v_{dip_i} + \sum_{l \in I_i} D_{kl, \Omega_i}  \mathcal{V}_{l} + \frac{1}{2} \mathcal{V}_{k} - \sum_{l \in I_i}S_{kl, \Omega_i} \vec{n}\cdot \bar{\bar{\sigma}}_i \nabla  \mathcal{V}_{l}.
\end{equation}
where to simplify the notations we introduced $v_{dip_j}(r) = \int_{ \Omega_j} G_j(r,r') f(r')dr'$ the potential due to the source in $\Omega_j$. Finally, enforcing the boundary condition \eqref{Cont} by subtracting \eqref{eq:Gamma0RprThOpJ} to  \eqref{eq:Gamma0RprThOp} we obtain
\begin{equation}\label{eq:FirstEqSym}
\begin{split}
\sum_{l \in I_i} D_{kl, \Omega_i} \mathcal{V}_{l} -\sum_{l \in \omega_i} D_{kl, \Omega_j} \mathcal{V}_{l} &+ \sum_{l \in I_i}S_{kl, \Omega_i} \vec{n}\cdot \bar{\bar{\sigma}}_i \nabla  \mathcal{V}_{l} \\
& - \sum_{l \in I_j}S_{kl, \Omega_j} \vec{n}\cdot \bar{\bar{\sigma}}_j \nabla  \mathcal{V}_{l} = -\left(\gamma_{0_{k}}^-v_{dip_i} - \gamma_{0_{k}}^+v_{dip_j}\right)
\end{split}
\end{equation}
The same reasoning using the trace of the conormal derivative gives the two following interior problems in $\Omega_j$ and $\Omega_i$
\begin{equation}\label{eq:Gamma1RprThOpJ}
\gamma_{1_{k}}^-V_i = \gamma_{1_{k}}^-v_{dip_i} + \sum_{l \in I_i} N_{kl, \Omega_i} \mathcal{V}_{l} + \frac{1}{2}\vec{n}\cdot \bar{\bar{\sigma}}_i \mathcal{V}_{k} - \sum_{l \in I_i}D^*_{kl, \Omega_i} \vec{n}\cdot \bar{\bar{\sigma}}_i \nabla \mathcal{V}_{l}
\end{equation}
and
\begin{equation}\label{eq:Gamma1RprThOp}
\gamma_{1_{k}}^+V_j = \gamma_{1_{k}}^+v_{dip_j} + \sum_{l \in I_j} N_{kl, \Omega_j} \mathcal{V}_{l} + \frac{1}{2}\vec{n}\cdot \bar{\bar{\sigma}}_j \mathcal{V}_{k} - \sum_{l \in I_j}D^*_{kl, \Omega_j} \vec{n}\cdot \bar{\bar{\sigma}}_j \nabla \mathcal{V}_{l}.
\end{equation}
Given that at the interface $\Gamma_{k}$, $\frac{1}{2}\vec{n}\cdot \bar{\bar{\sigma}}_i \mathcal{V}_{k} = \frac{1}{2}\vec{n}\cdot \bar{\bar{\sigma}}_j \mathcal{V}_{k}$, the equations \eqref{eq:Gamma1RprThOpJ} and \eqref{eq:Gamma1RprThOp} can be combined using the second boundary condition \eqref{ContDer} to get a second integral equation
\begin{equation}\label{eq:SecEqSym}
\begin{split}
\sum_{l \in I_i} N_{kl, \Omega_i} \mathcal{V}_{l} - \sum_{l \in I_j} N_{kl, \Omega_j} \mathcal{V}_{l} &+ \sum_{l \in I_i}D^*_{kl, \Omega_i} \vec{n}\cdot \bar{\bar{\sigma}}_i \nabla \mathcal{V}_{l}\\& - \sum_{l \in I_j}D^*_{kl, \Omega_j} \vec{n}\cdot \bar{\bar{\sigma}}_j \nabla \mathcal{V}_{l} = -\left(\gamma_{1_{k}}^-v_{dip_i} - \gamma_{1_{k}}^+v_{dip_j} \right) .
\end{split}
\end{equation}
The equations \eqref{eq:FirstEqSym} and \eqref{eq:SecEqSym} correspond to the symmetric formulation presented in \cite{kybic2005common}. However, the introduced formulation differs from it in its essence as the kernel of the involved integral operators is not the same. Moreover, the presenting reasoning generalizes the use of this formulation to fully inhomogeneous and anisotropic domains. As in the case of the standard symmetric formulation, at the external interface(s), only the boundary condition \eqref{ContDer} is necessary to be enforced, this means that for this(ese) interface(s), only \eqref{eq:SecEqSym} will be enforced. Indeed, \eqref{ContDer} at the outermost interface directly tells that the conormal derivative of the potential is zero here. }

\section{Discretization and Implementation of the Modified Symmetric Equation} \label{Discr}

To obtain a linear system to be solved, following the usual BEM strategy 
\cite{sauter2011boundary},
the integral equations must be tested with a suitably chosen set of basis functions. Tessellating the geometry into $N_t$ triangular cells $t_i$  and $N_v$ vertices $ v_i$, this gives rise to the linear system
\begin{equation}
	\mathbf{Z} \mathbf{a}=\mathbf{b}
\end{equation}
where $\mathbf{Z} $ contains the discrete version of the integral equation. Each entry of $\mathbf{Z} $ is  given by
\begin{equation}\label{eq:genBEM}
	\left[\mathbf{Z}\right]_{kl} = \int_{\mu^*_k}  {f_t}_{k}(r') Z{f_e}_{l}(r') dr'   
\end{equation}
where $Z$  is the  operator of the integral equation under consideration,  $\{f_{t_k}\}$ is the set of testing functions, and  $\{f_{e_k}\}$ is the set of expansion functions whose supports are given by  $\{\mu_{e_k}\}$. In our case, the unknowns \textcolor{black}{$\mathcal{V}_{i}$ and $\vec{n}\cdot \bar{\bar{\sigma_j}} \nabla \mathcal{V}_{i}$ can be respectively} discretized using, as expansion functions, piecewise linear functions $P_{1k} \in H^{1/2}$ and  piecewise constant functions $P_{0k}\in H^{-1/2}$. Such that,
\textcolor{black}{
	$\mathcal{V}_{j} \approx \sum_{k} \alpha_{j,k} P_{1k}$ and
$\bar{\bar{\sigma _i}} \nabla \mathcal{V}_{j} \approx \sum_{k} \beta_{j,k} P_{0k}$,
}
where $P_{0i}(r) = 1$ if $r\in t_i$ and $0$ elsewhere and $P_{1i}(r) = 1$ if $r=v_i$ and linearly decreases to $0$ on their support $\mu_{P_{1i}} = \{t_k|v_i \textnormal{ is a vertex of } t_k\}$.  
The support of $P_{0i}$ is given by $\mu_{P_{0i}} = t_i$. 

Expansion and testing functions should comply with the operatorial mappings of the involved integral operators. The range   of $S_{ij}$ \textcolor{black}{ and $D_{ij}$  is  $H^{1/2}$ for which the dual space is $H^{-1/2}$}. As $P_0 \in H^{-1/2}$ these functions can be used for testing  equation \textcolor{black}{ \eqref{eq:FirstEqSym}}. \textcolor{black}{ The operators involved in \eqref{eq:SecEqSym} are $D_{ij}^*$ and $N_{ij}$  whose range is $H^{-1/2}$. This means that pyramidal functions $P_{1} \in H^{1/2}$ are suitable testing functions. This discretization corresponds to the one presented in \cite{kybic2005common}. Then the entry of the matrix $\mathbf{Z}$ will be given by the discrete version of the operators $S_{ij,\Omega_k}, D_{ij,\Omega_k}, D^*_{ij,\Omega_k}$ and $N_{ij,\Omega_k}$ with
	\begin{subequations}
	\begin{align*}
	&\left[\mathbf{S_{ij,\Omega_k}}\right]_{kl} = \int_{\mu_{P_{0k}}} P_{0k} S_{ij,\Omega_k} P_{0l}(r)dr \\ 
	&\left[\mathbf{D_{ij,\Omega_k}}\right]_{kl} = \int_{\mu_{P_{0k}}} P_{0k} D_{ij,\Omega_k} P_{1l}(r)dr\\
	&\left[\mathbf{D^*_{ij,\Omega_k}}\right]_{kl} = \int_{\mu_{P_{1k}}} P_{1k} D^*_{ij,\Omega_k} P_{0l}(r)dr\\ 
	&\left[\mathbf{N_{ij,\Omega_k}}\right]_{kl} = \int_{\mu_{P_{1k}}} P_{1k} N_{ij,\Omega_k} P_{1l}(r)dr.
	\end{align*}
	\end{subequations}}

 \textcolor{black}{Note here that only surface basis functions are required. As such, only a surface mesh is necessary to implement the presented formulation. The separation of the considered geometry into different homogeneous anisotropic or isotropic compartments is however a necessary pre-processing step.}
 
\begin{figure}[!h]
	\centering
	\includegraphics[width=4cm]{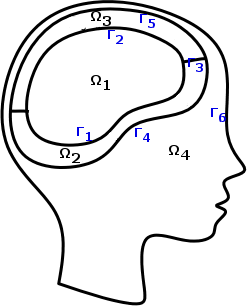}
	\caption{Topology of the domains used to write the exemplificatory system matrix in Section \ref{Discr}.}	\label{fig:SketchNonNestedSkull}
\end{figure}

\textcolor{black}{The proposed formulation can be interesting to model the skull anisotropy where in this layer, different conductivity tensors would correspond to different domains. 
Following this, as well as to fix the ideas and to simplify the implementation, we provide below the explicit form of the matrices in the case of the geometry depicted figure \ref{fig:SketchNonNestedSkull}. In this simplified case, the system matrix reads
\begin{equation*}
\mathbf{Z} =
\left[
\begin{matrix}
		\mathbf{S}& \mathbf{D}\\
		\mathbf{D^*}& \mathbf{N}
\end{matrix}
\right],
\end{equation*}	
with
\tiny
\begin{equation*}
\mathbf{A} = 	\left(
\begin{matrix}
\mathbf{A_{11,\Omega_1}} -\mathbf{A_{11,\Omega_2}} &  \mathbf{A_{12,\Omega_1}} & -\mathbf{A_{13,\Omega_2}} & -\mathbf{A_{14,\Omega_2}} & 0 & 0\\
\mathbf{A_{21, \Omega_1}} & \mathbf{A_{22, \Omega_1}} -\mathbf{A_{22, \Omega_3}} & -\mathbf{A_{23, \Omega_3}} & 0 & -\mathbf{A_{25, \Omega_3}} & 0\\
\mathbf{A_{31, \Omega_2}} & -\mathbf{A_{32, \Omega_3}} & \mathbf{A_{33, \Omega_2}} -\mathbf{A_{33, \Omega_3}}&  \mathbf{A_{34, \Omega_2}} & -\mathbf{A_{35, \Omega_3}} & 0\\
\mathbf{A_{41, \Omega_2}} &0 & \mathbf{A_{43, \Omega_2}}&  \mathbf{A_{44, \Omega_2}} -\mathbf{A_{44, \Omega_4}} & -\mathbf{A_{45, \Omega_4}} & -\mathbf{A_{46, \Omega_4}} \\
0 &\mathbf{A_{52, \Omega_3}} & \mathbf{A_{53, \Omega_3}} & -\mathbf{A_{54, \Omega_4}} &\mathbf{A_{55, \Omega_3}}- \mathbf{A_{55, \Omega_4}} & -\mathbf{A_{56, \Omega_4}} \\
0 & 0 & 0 &  \mathbf{A_{64, \Omega_4}} & \mathbf{A_{65, \Omega_4}} & \mathbf{A_{66, \Omega_4}} \\
\end{matrix}
\right),
\end{equation*}
\normalsize
where $\mathbf{A}$ stands for the operator $\mathbf{S}$, $\mathbf{D}$, $\mathbf{D^*}$ or $\mathbf{N}$. Note that the last line of $\mathbf{A}$ must not be computed when $\mathbf{A} = \mathbf{S}$ or $\mathbf{D}$, nor its last column when $\mathbf{A} = \mathbf{S}$ or $\mathbf{D^*}$. Indeed, equation \eqref{eq:FirstEqSym} is not necessary at the outermost interface and \eqref{ContDer} at the outermost interface directly gives $\vec{n}\cdot \bar{\bar{\sigma _4}}\nabla V_{4_6} = 0$.}

\textcolor{black}{The right hand side $\mathbf{b} = \left[\mathbf{b_{\gamma_0}} \, \mathbf{b_{\gamma_1}} \right]^T$ is defined using $\mathbf{b_{\gamma_p}} = \left[\mathbf{b_{\gamma_p 1}} \mathbf{b_{\gamma_p 2}},..., \mathbf{b_{\gamma_p N}}\right]^T $ with $p=0$ or $1$, $N=5$ if $p=0$ or $N=6$ if $p=1$, and
\begin{equation}
\left[\mathbf{b_{\gamma_p i}}\right]_{k} = \int_{\mu_k}  {f_t}_{k}(r') b_{\gamma_p i}(r') dr',
\end{equation}
where $b_{\gamma_p i}(r) = -\left(\gamma_{p_{i}}^-v_{dip_k}(r) - \gamma_{p_{i}}^+v_{dip_l}(r)\right)$, and ${f_t}_{k} = P_{1k}$ if $p=0$ and  ${f_t}_{k} = P_{0k}$ if $p=1$.  
In the same fashion, the vector $\mathbf{a}$ that contains the coefficients in the unknown expansion reads $\mathbf{a} = \left[ \mathbf{a_0} \, \mathbf{a_1}\right]^T$ with $\mathbf{a_p} = \left[\mathbf{a_{p 1}}, \mathbf{a_{p 2}},..., \mathbf{a_{p N}}\right]^T$ with $p=0$ or $1$, $N=5$ if $p=0$ or $N=6$ if $p=1$, and
\begin{equation}
\left\{
	\begin{split}
	&\left[\mathbf{a_{0j}}\right]_{l} = \beta_{j,l} \\
	&\left[\mathbf{a_{1j}}\right]_{l} = \alpha_{j,l}
	\end{split}
	\right.
\end{equation}
}

\subsection{\textcolor{black}{Computation of the innermost integral}}
In practice, computing the elements $[\mathbf{A_i}]_{kl} $ requires two integrations. The inner integral requires to compute either the integral of $G_j$ or of its derivative times the expansion function $f_{el}$. An analytical solution to obtain this integration for the usual Green's function $G(r,r') = \frac{1}{4\pi}\frac{1}{||r-r'||}$ has been proposed in \cite{graglia1993numerical}. In \eqref{GreenFct},  the change of variable $R = \sqrt{\bar{\bar \sigma}_i^{-1}}r$ (${\bar{\bar\sigma}}_i$ is a symmetric positive definite tensor) transforms $G_i(r,r')$ into $G_i(R,R') = \frac{1}{4\pi}\frac{1}{||R-R'||}$ for which we can apply the analytical integration formulas in \cite{graglia1993numerical}. The outer integrals  (arising from the testing of the operators) are performed numerically using Gaussian integration rules.

\subsection{\textcolor{black}{Computation of the hypersingular operator}}
Using the following representation \cite{sauter2011boundary}
\begin{equation}
\begin{split}
\int_{\mu_{P_{1m}}} & P_{1m}(r) N_{ij,\Omega_k}P_{1n}(r) d r \\
& = \int_{\mu_{P_{1m}} \times \mu_{P_{1n}}} G_k(r,r') \left(\bar{\bar{\sigma_k}}^{1/2} \nabla P_{1m}(r) \times \bar{\bar{\sigma_k}}^{1/2} \vec{n}\right)  \cdot \left(\bar{\bar{\sigma_k}}^{1/2} \nabla P_{1n}(r') \times \bar{\bar{\sigma_k}}^{1/2} \vec{n'} \right)dr dr'\\
& = \int_{\mu_{P_{1m}} \times \mu_{P_{1n}}} G_k(r,r') \det(\bar{\bar{\sigma_k}}) \left(\bar{\bar{\sigma_k}}^{\,-1} \nabla P_{1m}(r) \times \vec{n}\right)  \cdot \left( \nabla P_{1n}(r') \times  \vec{n'} \right)dr dr',
\end{split}
\end{equation}
 $\mathbf{N_{ij,\Omega_k}}$ can be implemented using the computation of $\mathbf{S_{ij,\Omega_k}}$.

\subsection{\textcolor{black}
{Application of the proposed method to volume meshes}}\label{subsecVolMesh}

A natural extension of the proposed method is to apply it to volume meshes. In this case, each tetrahedron can be considered as a unique domain with a given conductivity tensor. In regions where the conductivity is isotropic, tetrahedrons with equal  conductivities can be merged. This latter property, following from the fact that the formulation we are proposing is a surfacic one in nature, will warranty substantial savings in terms of the number of unknowns with respect to a standard volumetric method. The surface of the isotropic and homogeneous region is then given by the triangles separating tetrahedron with different conductivity tensors. In practice, if a volume mesh is considered, this requires an additional preprocessing step to extract the interfaces between regions with different conductivity tensors so that the considered geometry is a surfacic mesh. This can be done by comparing for each tetrahedron its conductivity tensor  with the conductivity tensor of its neighbor. The two tetrahedrons are merged  if the difference between their conductivity tensor is smaller than a chosen threshold $\epsilon$. For example, consider the two tetrahedrons $T_i$ and $T_j$ with conductivity tensors $\bar{\bar{\sigma}}_{i}$ and $\bar{\bar{\sigma}}_{j}$ respectively, and boundaries given by the set of triangles $\bigcup_{k} t_{ik}$ and $\bigcup_{k} t_{jk}$ respectively. The interface $t_{ij} = \Gamma_{ij}$ between $T_i$ and $T_j$ is taken into account only if $||\bar{\bar{\sigma}}_{i}- \bar{\bar{\sigma}}_{j}|| > \epsilon$, otherwise $T_i \cup T_j$ will form a unique domain $\Omega_i$ such that $\partial \Omega_i = \{\bigcup_{k} t_{ik} \cup \bigcup_{k} t_{jk}\} \setminus t_{ij}$. To treat the full volume mesh, the process is repeated for all the tetrahedrons. In the case where the mesh is fully inhomogeneous or when no tetrahedron at all are merged, the system matrix $\mathbf{Z}$ will be extremely sparse since for one line, only seven columns of the matrix are not zero. \textcolor{black}{The reader should keep in mind that using volume mesh is not mandatory to use the presented formulation. It is only a convenient pre-processing step in our situation since, moreover, comparison with volume methods (FEM) will be carried out. Indeed, this step is independent from the integral formulation itself and a pure surface mesh can directly be employed.
}\textcolor{black}{However, the correct modelling of anisotropic compartments requires to subdivide them. As a consequence, while isotropic compartments are modelled with surfaces, anisotropic ones are virtually using volume elements.}

\section{Numerical Results}\label{NumRes}

Two sets of numerical experiments have been carried out. Since in case of isotropic conductivity, the presented method is the same as the standard symmetric formulation, we focus on anisotropic conductivity profiles. The first set aimed at assessing the accuracy of the proposed method with canonical anisotropic head meshes (layered anisotropic spheres) while the second set aimed at showing its ability to handle realistic anisotropic conductivity profiles and meshes.

\subsection{\textcolor{black}{Modelling of the anisotropy with volume meshes}}\label{skullAnis}

In modelling the skull layer different approaches are found in the literature. The work of  \cite{akhtari2002conductivities} has shown that the different skull layers, a cancellous bone in between two layers of compact bone, \textcolor{black}{do not have the same conductivities}. 
\textcolor{black}{As a consequence, different possibilities are available to model this layer: one can model the skull with a simple isotropic layer, a single anisotropic layer or three layers with different isotropic conductivities. In \cite{dannhauer2011modeling}, the authors investigate how to model the skull layer. Their recommendation depends on the available skull geometry accuracy.}
Obtaining a precise mesh \textcolor{black}{to make out} the three layers is not without computational overload and it might explains why modeling the skull with a single anisotropic layer is still very common as for example in \cite{wolters2006influence}, \cite{gullmar2010influence} or \cite{acar2013effects}.
In this paragraph we present results on spherical layered meshes where the skull is modelled with a single anisotropic layer. The method, however, can of course be applied also in the case of simulations where the skull layer is subdivided into its three isotropic composing layers. The first experiments are based on a three layers sphere mesh of radii 0.87, 0.92 and 1. These layers stand for the brain, skull and  scalp respectively. In the first and last layer, the conductivity is set to $1$ while the conductivity in the middle layer is assumed to be radially anisotropic. The conductivity ratio between the first and the second layer is $\sigma_{skull,iso} = 1/15$ , following  \cite{oostendorp2000conductivity}. The ratio between the radial and tangential conductivity in the skull is $1/10$. The radial and tangential conductivity tensor component in each tetrahedra is computed using 
	\cite{wolters2006influence}
	\begin{equation}\label{condTensorEq}
	\frac{4}{3}\pi\sigma_{r}(\sigma_{t})^2 = \frac{4}{3}\pi\sigma_{skull,iso}^3.
	\end{equation}

\begin{figure}[h!]
	\centering
 	\subfigure[\textcolor{black}{Volume mesh} \label{fig:label:VolumeMesh}]{\includegraphics[width=0.2\textwidth]{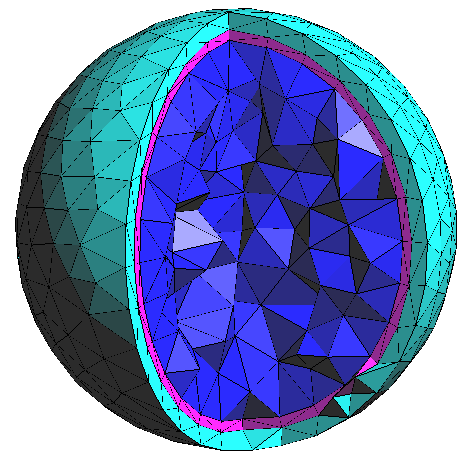}}
 	\subfigure[\textcolor{black}{Corresponding surface.}\label{fig:label:MergedMesh}]{\includegraphics[width=0.2\textwidth]{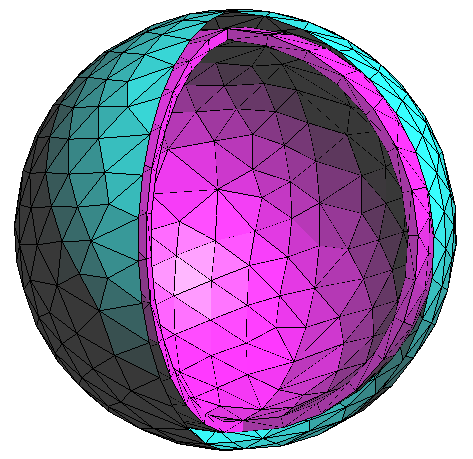}}
 	\caption{\textcolor{black}{Volume mesh before \protect\subref{fig:label:VolumeMesh} and after preprocessing, according to the conductivity value of the tetrahedra. \protect\subref{fig:label:MergedMesh}}}\label{fig:mergingVolumes}
\end{figure}

	 We begin with a volume mesh and extract a surface mesh by comparing the conductivity values in each tetrahedron following the process described in paragraph \ref{subsecVolMesh}. In this fashion, the tetrahedra in the brain and scalp layers are gathered to form only two homogeneous and isotropic domains while the tetrahedra of the skull layer form different domains depending on the accepted conductivity error $\epsilon$ between two neighboring tetrahedra. Figure~\ref{fig:mergingVolumes} shows the volume mesh and the surface mesh obtained.

\subsubsection{Convergence of the solution}  
\begin{figure}\label{anisSphe}
	\centering
	\includegraphics[width=0.7\textwidth]{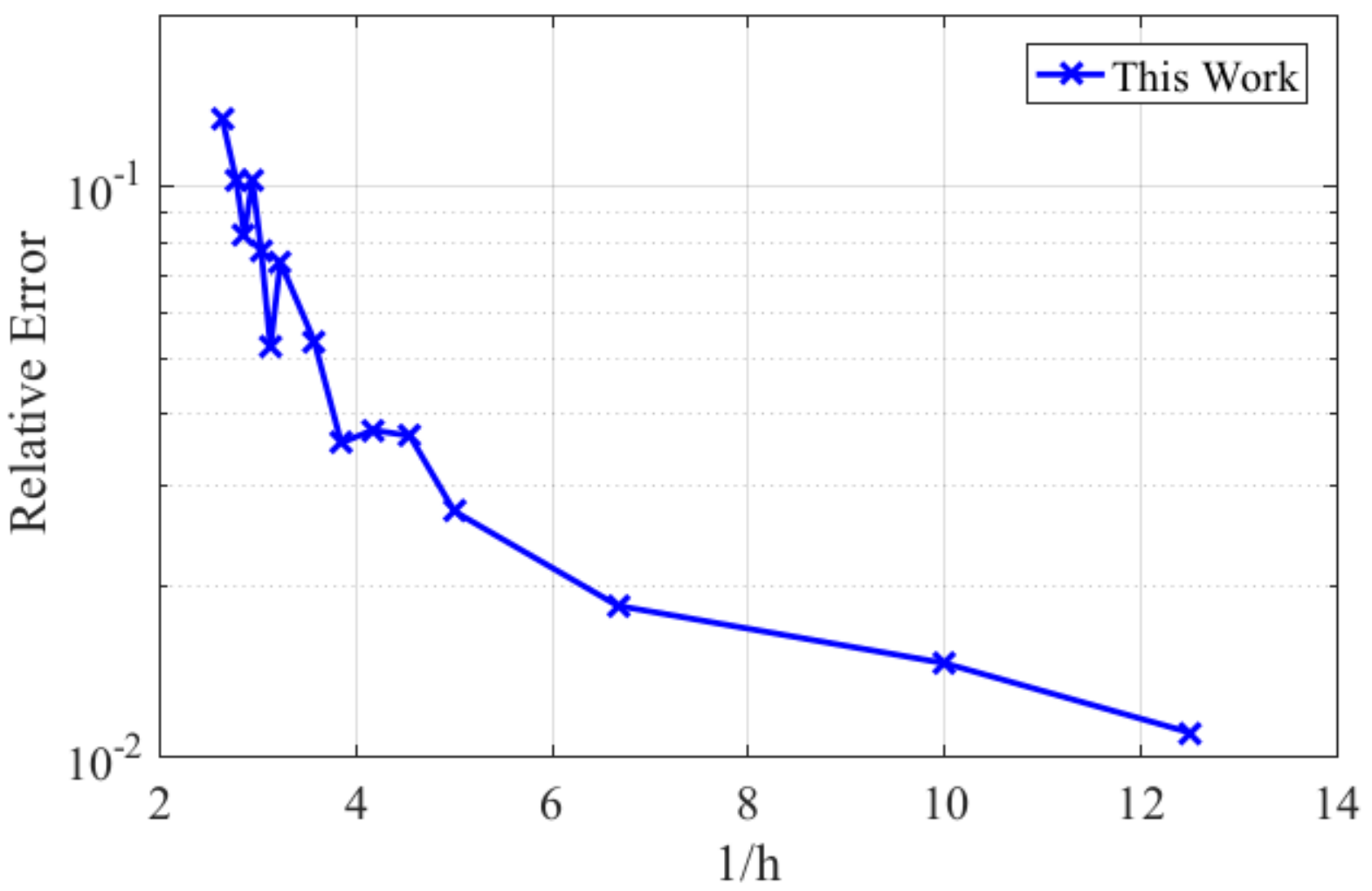}
	\caption{\textcolor{black}{Convergence of the solution of the proposed equation when increasing the number of unknowns.}
	}
	\label{fig:resAnis}
\end{figure}

In this first sets of experiment, $\epsilon$ was chosen small enough so that no tetrahedron of the skull layer is merged. This first experiment focused on verifying the convergence of the newly proposed formulation when the number of unknowns is increased.
\textcolor{black}{
The accuracy is assessed by calculating 
the relative error of the solution given by the proposed integral formulation 
with respect to the analytic reference solution 
available in \cite{zhang1995fast}. The mesh refinement parameter, denoted with $h$, was decreased from $0.38$ to $0.08$ so that the initial number of tetrahedra grew from $3607$ to $112827$. Applying the preprocessing described in \ref{subsecVolMesh} gave a number of domain increasing from $1495$ to $33117$.} A single dipole source with unitary moment along the z-axis was placed in the center of the innermost sphere\textcolor{black}{. The dipole position influence on the numerical results is studied in a second scenario}. The results for this test are shown in Figure~\ref{fig:resAnis}. It is clear that the relative error decreases, showing the convergence and the consistency of the proposed approach. \textcolor{black}{ The time that our prototype code needed to solve the anisotropic EEG forward \textcolor{black}{problem} in the presented canonical scenario is presented Table \ref{table:timings}.}


\begin{table}[h]
\centering
\small{
\textcolor{black}{
   \begin{tabular}{|c|c|c|c|c|}
   \hline
     Time Z & Time sol. & N elements & sparsity & Relative Error  \\
   \hline
    $545$ s & $12.7$ s & $19,392$ & $0.1$ & $0.013$\\
   \hline
\end{tabular}}}
\caption{\textcolor{black}{Timing results to obtain an error of $1\%$ with respect to the analytical anisotropic solution. "Time Z" is the time to build the system matrix,  "Time sol" is the time to solve the system (direct inversion), "N elements"is the number of basis functions, sparsity is the ratio of the number of non-zero of the matrix with respect to the size of the matrix.} \label{table:timings}}
\end{table}


\subsubsection{Accuracy for randomized dipole positions and different dipole eccentricities}
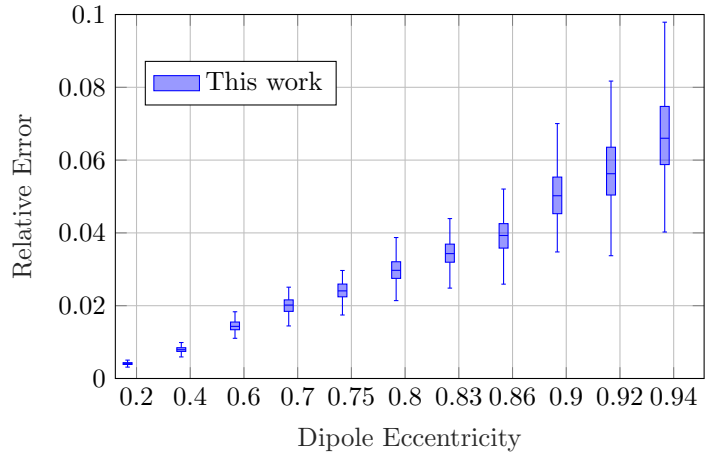
\begin{figure}
	\centering
	\input{RE.tex}
		\caption{\textcolor{black}{
				Relative error versus dipole eccentricity in the computation of 200 dipole source both for radial and tangential dipole orientation for the method proposed in this work.}
		}
		\label{fig:Random}
	\end{figure}

Since one dipole may not be representative of the complex brain electric sources pattern of activation, we carried out another experiment, with the same anisotropic conductivity parameters, in which dipole positions are randomly set. For each eccentricity, we computed the relative error in 200 simulations of 1 dipole source, for radial and tangential unitary dipole moment. \textcolor{black}{The obtained results, shown in Figure~\ref{fig:Random},  confirm that the proposed approach provides a formulation which is able to handle anisotropic conductivity profile for any dipole position.}

\subsection{Application to a realistic head mesh}

\begin{figure}[h!]
	\centering
	\subfigure[\textcolor{black}{Volume mesh.} \label{fig:label:RealVolMesh}]{\includegraphics[width=0.4\textwidth]{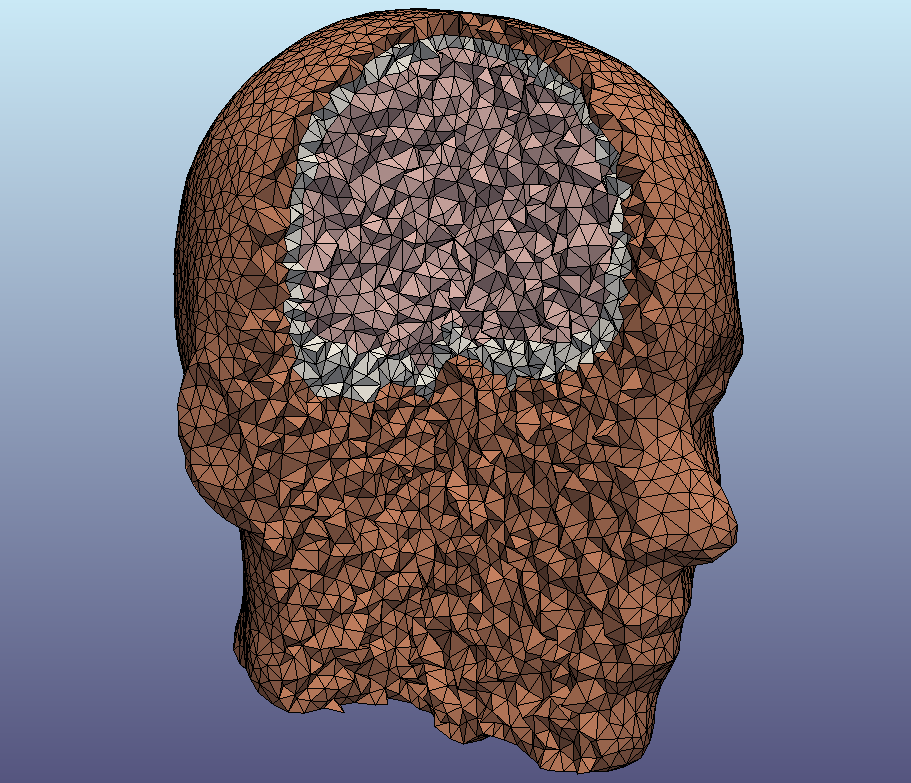}}
	\subfigure[\textcolor{black}{Extracted surface mesh.}\label{fig:label:RealSurfMesh}]{\includegraphics[width=0.4\textwidth]{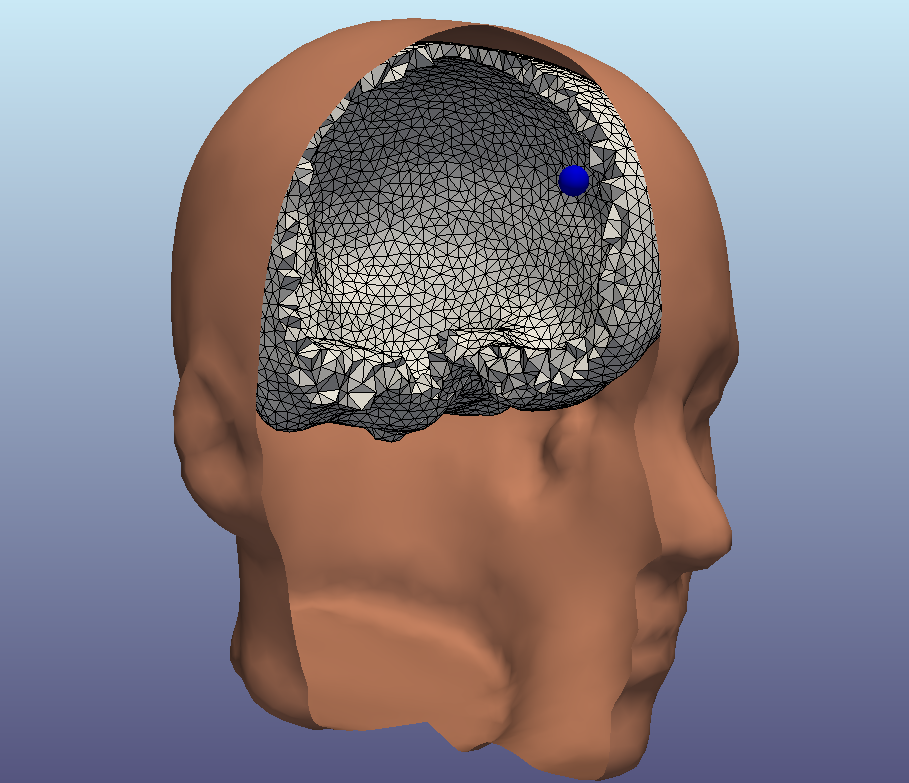}}
	\caption{\textcolor{black}{Three Layers mesh used in the realistic simulation : a volume mesh \protect\subref{fig:label:RealVolMesh} used in the FEM solution and the extracted surface mesh used with the proposed formulation \protect\subref{fig:label:RealSurfMesh}. The blue sphere on the right picture shows the position of the dipole used in the simulation.}}
\end{figure}

\begin{figure}[h!]
	\centering
	\includegraphics[width=0.35\textwidth]{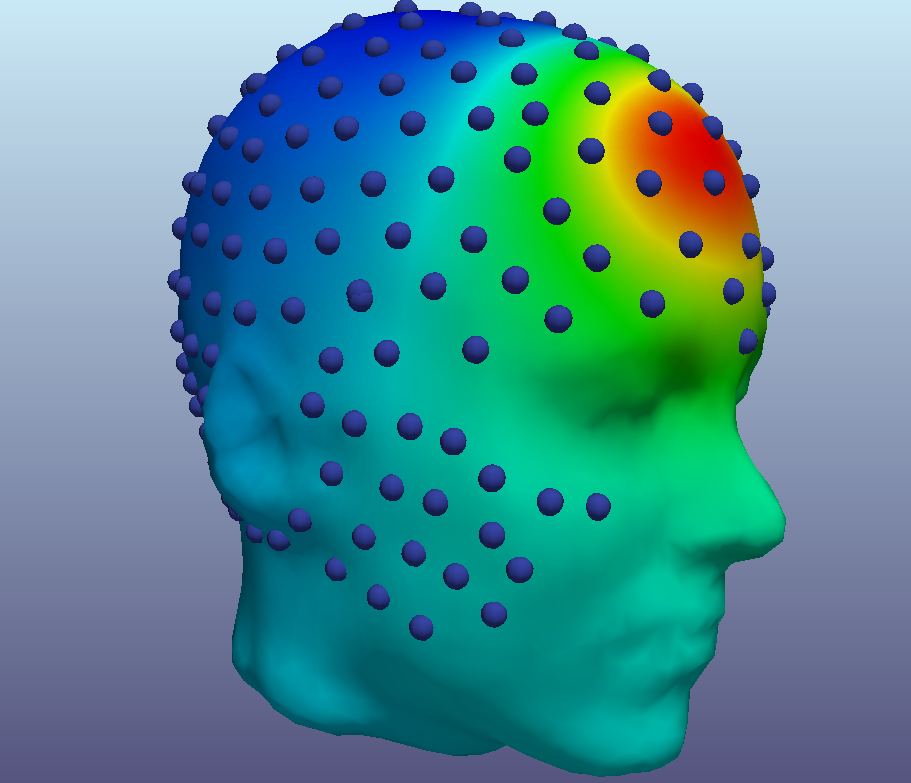}
	\caption{\textcolor{black}{Computed potential on the scalp of a realistic mesh.}     
	\label{fig:ScalpPotential}}
\end{figure}

\begin{figure}[h!]
	\centering
     \input{Electrodes_AtPotential.tex}
	\caption{\textcolor{black}{Potential computed at 256 electrodes positions.}}
	\label{fig:ElectrodesPotential}
\end{figure}
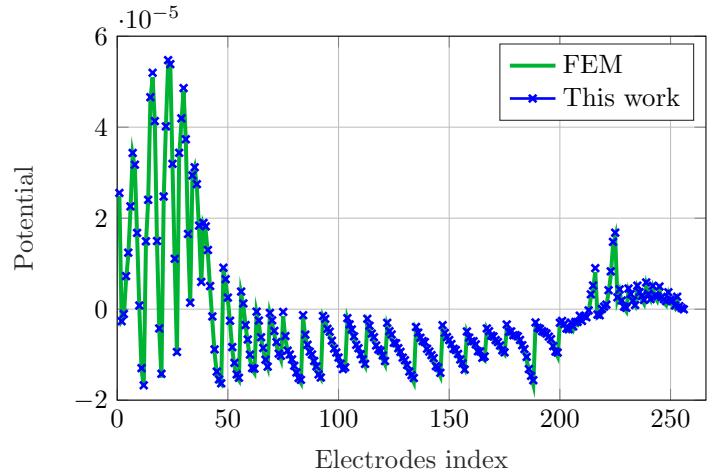

\normalsize
The second experiment  is the comparison of the potential obtained in a realistic case with the proposed method and the FEM formulation. Indeed, from grey scale data given by MRI, it is possible to extract volume meshes, (see for example \cite{Fang2009tetrahedra}), that are usually used in FEM simulations. From this volume mesh, our method can extract surfaces that delimit domains with the same conductivity. In the presented example, we used \cite{oostenveld2010fieldtrip} to generate a three layers volume head mesh. The fully tetrahedral mesh was used in the FEM solution. A conductivity tensor was associated to each tetrahedra following the choice of the previous anisotropic canonical example. Then, the preprocessing subroutine of our scheme provided a surface mesh for which we applied directly the proposed method. This mesh is shown figure~\ref{fig:label:RealSurfMesh}.
The number of tetrahedra was reduced from $187317$ to $51709$ triangular cells. The choice of a merging threshold $\epsilon = 0.001$ led to the generation of $11698$ domains. The obtained scalp potential is shown in figure~\ref{fig:ScalpPotential}. A relative error of $4.5\%$ between the potential computed with the proposed formulation and the fully anisotropic FEM solution was obtained. The potential computed at 256 electrodes positions is presented in figure~\ref{fig:ElectrodesPotential}. This figure shows the overlapping curves obtained with the two methods and confirms the validity of the approach.
\textcolor{black}{ This realistic scenario shows the ability of the proposed method to handle realistic meshes and in particular}, \textcolor{black}{if wanted}\textcolor{black}{, the skull anisotropy. }\textcolor{black}{In case white matter fiber anisotropy is included, no substantial changes must be done\textcolor{black}{:} no assumption was made on the geometry of the media. As a consequence, the only difference would reside in the obtention of the mesh. In practice, the number of degrees of freedom may require the use of a fast solver.}

\section{Discussion and Conclusion} \label{discCcl}

\textcolor{black}{In this paper, a new integral formulation for the EEG forward problem that takes into account anisotropic conductivity profiles was presented. The integral formulation is obtained by leveraging on the representation theorem using an anisotropic fundamental solution. Implementations details are provided and explain the discretization strategy. In particular, a strategy is proposed to tackle inhomogeneous and anisotropic conductivity profile with volume meshes. In this case, the level of inhomogeneity can be controlled by choosing an accepted error in the anisotropic conductivity when building a surfacic mesh from a volume mesh. The level of inhomogeneity does not decrease the accuracy of the solution, on the contrary, it helps to provide a better anatomical model. The numerical results show the correctness of the approach. 
\textcolor{black}{In the canonical cases, the results given by the new integral equation are matching the analytical solution. In the realistic case, the solution obtained with the proposed formulation is compatible with the one provided by the Finite Elements formulation.} One shortcoming of boundary integral formulations, when compared to FEM, is that the arising system matrix is fully populated while in the case of FEM the system matrix is sparse. 
However, in the presented formulation, as in the symmetric formulation, 
\textcolor{black}{the number of entries of the system matrix depends on the number of interfaces\textcolor{black}{:} the system matrix is block diagonal with the size of the blocks depending on the number of elements in the considered interface.} Furthermore, a popular way to fix this disadvantage is to resort to fast solving techniques such as Adaptive Cross Approximation (ACA) or Fast Multipole Method (FMM) that allows to build the system matrix in a linear complexity. 
}

	\section*{Acknowledgement}
This work was supported in part by the CominLabs excellence laboratory, managed by the National Research Agency in the “Investing for the Future” program under reference ANR-10-LABX-07-01, project SABRE and 
in part by the European Research Council (ERC) under the European Union’s Horizon 2020 research and innovation program 
(ERC project 321, grant No. 724846).

\section*{References}

\bibliography{BiblioAnisFull}

\end{document}

%% file: RE.tex
%
%
\definecolor{mycolor1}{rgb}{1,1,1}%
\begin{tikzpicture}

\begin{axis}[%
width=3.05in,
height=1.9in,
at={(2.275in,1.015in)},
scale only axis,
xmin=2.57921010638632,
xmax=10.8,
xtick={1.375,2.125,2.875,3.625,4.375,5.125,5.875,6.625,7.375,8.125,8.875,9.625,10.375,11.125,11.875},
xticklabels={{0.01},{0.1},{0.2},{0.4},{0.6},{0.7},{0.75},{0.8},{0.83},{0.86},{0.9},{0.92},{0.94},{0.96},{0.98}},
xlabel style={font=\color{white!15!black}},
xlabel={Dipole Eccentricity},
ymin=-0.0,
ymax=0.1,
ylabel style={font=\color{white!15!black}},
ylabel={Relative Error},
yticklabel style={
            /pgf/number format/fixed,
            /pgf/number format/precision=2
        },
axis background/.style={fill=white},
xmajorgrids,
ymajorgrids,
legend style={at={(0.05,0.75)}, anchor=south west, legend cell align=left, align=left, draw=black, fill=mycolor1}
]
\addplot [color=red, forget plot]
  table[row sep=crcr]{%
1.25	0.00246572168384106\\
1.25	0.00274260306465595\\
};

\addplot [color=blue, forget plot]
  table[row sep=crcr]{%
2	0.00305505939019079\\
2	0.00358002182550168\\
};

\addplot [color=blue, forget plot]
  table[row sep=crcr]{%
2.75	0.00434931820696281\\
2.75	0.00504963380276746\\
};

\addplot [color=blue, forget plot]
  table[row sep=crcr]{%
3.5	0.00846048125195737\\
3.5	0.0098826572904111\\
};

\addplot [color=blue, forget plot]
  table[row sep=crcr]{%
4.25	0.0155098354112695\\
4.25	0.0183329399035189\\
};

\addplot [color=blue, forget plot]
  table[row sep=crcr]{%
5	0.0215985909175807\\
5	0.0250876708798214\\
};

\addplot [color=blue, forget plot]
  table[row sep=crcr]{%
5.75	0.0259434246274605\\
5.75	0.0296763200836182\\
};

\addplot [color=blue, forget plot]
  table[row sep=crcr]{%
6.5	0.032076267173919\\
6.5	0.0387253940898203\\
};

\addplot [color=blue, forget plot]
  table[row sep=crcr]{%
7.25	0.0369130348133899\\
7.25	0.0439357203351363\\
};

\addplot [color=blue, forget plot]
  table[row sep=crcr]{%
8	0.0425473796080108\\
8	0.0520485838184162\\
};

\addplot [color=blue, forget plot]
  table[row sep=crcr]{%
8.75	0.0553339694794634\\
8.75	0.0700370338872719\\
};

\addplot [color=blue, forget plot]
  table[row sep=crcr]{%
9.5	0.0635385457962645\\
9.5	0.0817123278273342\\
};

\addplot [color=blue, forget plot]
  table[row sep=crcr]{%
10.25	0.0747470256490118\\
10.25	0.0978880774343725\\
};

\addplot [color=blue, forget plot]
  table[row sep=crcr]{%
11	0.0922120294729493\\
11	0.123979659610648\\
};

\addplot [color=blue, forget plot]
  table[row sep=crcr]{%
11.75	1.17100520093545\\
11.75	1.75171664770566\\
};

\addplot [color=blue, forget plot]
  table[row sep=crcr]{%
1.25	0.0019445718175763\\
1.25	0.00217805784732714\\
};

\addplot [color=blue, forget plot]
  table[row sep=crcr]{%
2	0.00214673503070858\\
2	0.00268616945419024\\
};

\addplot [color=blue, forget plot]
  table[row sep=crcr]{%
2.75	0.00312546514809393\\
2.75	0.00385834157296039\\
};

\addplot [color=blue, forget plot]
  table[row sep=crcr]{%
3.5	0.00592574316318846\\
3.5	0.00743810030067779\\
};

\addplot [color=blue, forget plot]
  table[row sep=crcr]{%
4.25	0.0110436853813222\\
4.25	0.0133985955129327\\
};

\addplot [color=blue, forget plot]
  table[row sep=crcr]{%
5	0.0144381045907011\\
5	0.018445177141226\\
};

\addplot [color=blue, forget plot]
  table[row sep=crcr]{%
5.75	0.0174445228622804\\
5.75	0.0224406487794505\\
};

\addplot [color=blue, forget plot]
  table[row sep=crcr]{%
6.5	0.0213965994759996\\
6.5	0.027506167645826\\
};

\addplot [color=blue, forget plot]
  table[row sep=crcr]{%
7.25	0.0248307018092983\\
7.25	0.0319482624363688\\
};

\addplot [color=blue, forget plot]
  table[row sep=crcr]{%
8	0.0259191792219182\\
8	0.0358210565015841\\
};

\addplot [color=blue, forget plot]
  table[row sep=crcr]{%
8.75	0.0347671322944235\\
8.75	0.0452818983615675\\
};

\addplot [color=blue, forget plot]
  table[row sep=crcr]{%
9.5	0.0337327439276514\\
9.5	0.0504091583528547\\
};

\addplot [color=blue, forget plot]
  table[row sep=crcr]{%
10.25	0.0402441220709318\\
10.25	0.0587702634372203\\
};

\addplot [color=blue, forget plot]
  table[row sep=crcr]{%
11	0.0483254930222444\\
11	0.0691240306141183\\
};

\addplot [color=blue, forget plot]
  table[row sep=crcr]{%
11.75	0.0493200829125501\\
11.75	0.101232008042612\\
};

\addplot [color=blue, forget plot]
  table[row sep=crcr]{%
1.21875	0.00274260306465595\\
1.28125	0.00274260306465595\\
};

\addplot [color=blue, forget plot]
  table[row sep=crcr]{%
1.96875	0.00358002182550168\\
2.03125	0.00358002182550168\\
};

\addplot [color=blue, forget plot]
  table[row sep=crcr]{%
2.71875	0.00504963380276746\\
2.78125	0.00504963380276746\\
};

\addplot [color=blue, forget plot]
  table[row sep=crcr]{%
3.46875	0.0098826572904111\\
3.53125	0.0098826572904111\\
};

\addplot [color=blue, forget plot]
  table[row sep=crcr]{%
4.21875	0.0183329399035189\\
4.28125	0.0183329399035189\\
};

\addplot [color=blue, forget plot]
  table[row sep=crcr]{%
4.96875	0.0250876708798214\\
5.03125	0.0250876708798214\\
};

\addplot [color=blue, forget plot]
  table[row sep=crcr]{%
5.71875	0.0296763200836182\\
5.78125	0.0296763200836182\\
};

\addplot [color=blue, forget plot]
  table[row sep=crcr]{%
6.46875	0.0387253940898203\\
6.53125	0.0387253940898203\\
};

\addplot [color=blue, forget plot]
  table[row sep=crcr]{%
7.21875	0.0439357203351363\\
7.28125	0.0439357203351363\\
};

\addplot [color=blue, forget plot]
  table[row sep=crcr]{%
7.96875	0.0520485838184162\\
8.03125	0.0520485838184162\\
};

\addplot [color=blue, forget plot]
  table[row sep=crcr]{%
8.71875	0.0700370338872719\\
8.78125	0.0700370338872719\\
};

\addplot [color=blue, forget plot]
  table[row sep=crcr]{%
9.46875	0.0817123278273342\\
9.53125	0.0817123278273342\\
};

\addplot [color=blue, forget plot]
  table[row sep=crcr]{%
10.21875	0.0978880774343725\\
10.28125	0.0978880774343725\\
};

\addplot [color=blue, forget plot]
  table[row sep=crcr]{%
10.96875	0.123979659610648\\
11.03125	0.123979659610648\\
};

\addplot [color=blue, forget plot]
  table[row sep=crcr]{%
11.71875	1.75171664770566\\
11.78125	1.75171664770566\\
};

\addplot [color=blue, forget plot]
  table[row sep=crcr]{%
1.21875	0.0019445718175763\\
1.28125	0.0019445718175763\\
};

\addplot [color=blue, forget plot]
  table[row sep=crcr]{%
1.96875	0.00214673503070858\\
2.03125	0.00214673503070858\\
};

\addplot [color=blue, forget plot]
  table[row sep=crcr]{%
2.71875	0.00312546514809393\\
2.78125	0.00312546514809393\\
};

\addplot [color=blue, forget plot]
  table[row sep=crcr]{%
3.46875	0.00592574316318846\\
3.53125	0.00592574316318846\\
};

\addplot [color=blue, forget plot]
  table[row sep=crcr]{%
4.21875	0.0110436853813222\\
4.28125	0.0110436853813222\\
};

\addplot [color=blue, forget plot]
  table[row sep=crcr]{%
4.96875	0.0144381045907011\\
5.03125	0.0144381045907011\\
};

\addplot [color=blue, forget plot]
  table[row sep=crcr]{%
5.71875	0.0174445228622804\\
5.78125	0.0174445228622804\\
};

\addplot [color=blue, forget plot]
  table[row sep=crcr]{%
6.46875	0.0213965994759996\\
6.53125	0.0213965994759996\\
};

\addplot [color=blue, forget plot]
  table[row sep=crcr]{%
7.21875	0.0248307018092983\\
7.28125	0.0248307018092983\\
};

\addplot [color=blue, forget plot]
  table[row sep=crcr]{%
7.96875	0.0259191792219182\\
8.03125	0.0259191792219182\\
};

\addplot [color=blue, forget plot]
  table[row sep=crcr]{%
8.71875	0.0347671322944235\\
8.78125	0.0347671322944235\\
};

\addplot [color=blue, forget plot]
  table[row sep=crcr]{%
9.46875	0.0337327439276514\\
9.53125	0.0337327439276514\\
};

\addplot [color=blue, forget plot]
  table[row sep=crcr]{%
10.21875	0.0402441220709318\\
10.28125	0.0402441220709318\\
};

\addplot [color=blue, forget plot]
  table[row sep=crcr]{%
10.96875	0.0483254930222444\\
11.03125	0.0483254930222444\\
};

\addplot [color=blue, forget plot]
  table[row sep=crcr]{%
11.71875	0.0493200829125501\\
11.78125	0.0493200829125501\\
};

\addplot [color=blue, forget plot]
  table[row sep=crcr]{%
1.1875	0.00227844623754188\\
1.3125	0.00227844623754188\\
};

\addplot [color=blue, forget plot]
  table[row sep=crcr]{%
1.9375	0.00287834488404466\\
2.0625	0.00287834488404466\\
};

\addplot [color=blue, forget plot]
  table[row sep=crcr]{%
2.6875	0.00407700690995692\\
2.8125	0.00407700690995692\\
};

\addplot [color=blue, forget plot]
  table[row sep=crcr]{%
3.4375	0.00791030406685853\\
3.5625	0.00791030406685853\\
};

\addplot [color=blue, forget plot]
  table[row sep=crcr]{%
4.1875	0.0143198456386902\\
4.3125	0.0143198456386902\\
};

\addplot [color=blue, forget plot]
  table[row sep=crcr]{%
4.9375	0.0201752175368197\\
5.0625	0.0201752175368197\\
};

\addplot [color=blue, forget plot]
  table[row sep=crcr]{%
5.6875	0.0240725356094549\\
5.8125	0.0240725356094549\\
};

\addplot [color=blue, forget plot]
  table[row sep=crcr]{%
6.4375	0.0297150399768907\\
6.5625	0.0297150399768907\\
};

\addplot [color=blue, forget plot]
  table[row sep=crcr]{%
7.1875	0.0343110942232504\\
7.3125	0.0343110942232504\\
};

\addplot [color=blue, forget plot]
  table[row sep=crcr]{%
7.9375	0.0392867798966865\\
8.0625	0.0392867798966865\\
};

\addplot [color=blue, forget plot]
  table[row sep=crcr]{%
8.6875	0.0502087180384684\\
8.8125	0.0502087180384684\\
};

\addplot [color=blue, forget plot]
  table[row sep=crcr]{%
9.4375	0.0562609335850682\\
9.5625	0.0562609335850682\\
};

\addplot [color=blue, forget plot]
  table[row sep=crcr]{%
10.1875	0.066004462230111\\
10.3125	0.066004462230111\\
};

\addplot [color=blue, forget plot]
  table[row sep=crcr]{%
10.9375	0.0800857530645725\\
11.0625	0.0800857530645725\\
};

\addplot [color=blue, forget plot]
  table[row sep=crcr]{%
11.6875	0.148796398294602\\
11.8125	0.148796398294602\\
};
;


\addplot[area legend, draw=blue, fill=blue, fill opacity=0.4]
table[row sep=crcr] {%
x	y\\
11.6875	0.101232008042612\\
11.6875	1.17100520093545\\
11.8125	1.17100520093545\\
11.8125	0.101232008042612\\
11.6875	0.101232008042612\\
}--cycle;
\addlegendentry{This work}

\addplot[area legend, draw=blue, fill=blue, fill opacity=0.4, forget plot]
table[row sep=crcr] {%
x	y\\
10.9375	0.0691240306141183\\
10.9375	0.0922120294729493\\
11.0625	0.0922120294729493\\
11.0625	0.0691240306141183\\
10.9375	0.0691240306141183\\
}--cycle;

\addplot[area legend, draw=blue, fill=blue, fill opacity=0.4, forget plot]
table[row sep=crcr] {%
x	y\\
10.1875	0.0587702634372203\\
10.1875	0.0747470256490118\\
10.3125	0.0747470256490118\\
10.3125	0.0587702634372203\\
10.1875	0.0587702634372203\\
}--cycle;

\addplot[area legend, draw=blue, fill=blue, fill opacity=0.4, forget plot]
table[row sep=crcr] {%
x	y\\
9.4375	0.0504091583528547\\
9.4375	0.0635385457962645\\
9.5625	0.0635385457962645\\
9.5625	0.0504091583528547\\
9.4375	0.0504091583528547\\
}--cycle;

\addplot[area legend, draw=blue, fill=blue, fill opacity=0.4, forget plot]
table[row sep=crcr] {%
x	y\\
8.6875	0.0452818983615675\\
8.6875	0.0553339694794634\\
8.8125	0.0553339694794634\\
8.8125	0.0452818983615675\\
8.6875	0.0452818983615675\\
}--cycle;

\addplot[area legend, draw=blue, fill=blue, fill opacity=0.4, forget plot]
table[row sep=crcr] {%
x	y\\
7.9375	0.0358210565015841\\
7.9375	0.0425473796080108\\
8.0625	0.0425473796080108\\
8.0625	0.0358210565015841\\
7.9375	0.0358210565015841\\
}--cycle;

\addplot[area legend, draw=blue, fill=blue, fill opacity=0.4, forget plot]
table[row sep=crcr] {%
x	y\\
7.1875	0.0319482624363688\\
7.1875	0.0369130348133899\\
7.3125	0.0369130348133899\\
7.3125	0.0319482624363688\\
7.1875	0.0319482624363688\\
}--cycle;

\addplot[area legend, draw=blue, fill=blue, fill opacity=0.4, forget plot]
table[row sep=crcr] {%
x	y\\
6.4375	0.027506167645826\\
6.4375	0.032076267173919\\
6.5625	0.032076267173919\\
6.5625	0.027506167645826\\
6.4375	0.027506167645826\\
}--cycle;

\addplot[area legend, draw=blue, fill=blue, fill opacity=0.4, forget plot]
table[row sep=crcr] {%
x	y\\
5.6875	0.0224406487794505\\
5.6875	0.0259434246274605\\
5.8125	0.0259434246274605\\
5.8125	0.0224406487794505\\
5.6875	0.0224406487794505\\
}--cycle;

\addplot[area legend, draw=blue, fill=blue, fill opacity=0.4, forget plot]
table[row sep=crcr] {%
x	y\\
4.9375	0.018445177141226\\
4.9375	0.0215985909175807\\
5.0625	0.0215985909175807\\
5.0625	0.018445177141226\\
4.9375	0.018445177141226\\
}--cycle;

\addplot[area legend, draw=blue, fill=blue, fill opacity=0.4, forget plot]
table[row sep=crcr] {%
x	y\\
4.1875	0.0133985955129327\\
4.1875	0.0155098354112695\\
4.3125	0.0155098354112695\\
4.3125	0.0133985955129327\\
4.1875	0.0133985955129327\\
}--cycle;

\addplot[area legend, draw=blue, fill=blue, fill opacity=0.4, forget plot]
table[row sep=crcr] {%
x	y\\
3.4375	0.00743810030067779\\
3.4375	0.00846048125195737\\
3.5625	0.00846048125195737\\
3.5625	0.00743810030067779\\
3.4375	0.00743810030067779\\
}--cycle;

\addplot[area legend, draw=blue, fill=blue, fill opacity=0.4, forget plot]
table[row sep=crcr] {%
x	y\\
2.6875	0.00385834157296039\\
2.6875	0.00434931820696281\\
2.8125	0.00434931820696281\\
2.8125	0.00385834157296039\\
2.6875	0.00385834157296039\\
}--cycle;

\addplot[area legend, draw=blue, fill=blue, fill opacity=0.4, forget plot]
table[row sep=crcr] {%
x	y\\
1.9375	0.00268616945419024\\
1.9375	0.00305505939019079\\
2.0625	0.00305505939019079\\
2.0625	0.00268616945419024\\
1.9375	0.00268616945419024\\
}--cycle;

\addplot[area legend, draw=blue, fill=blue, fill opacity=0.4, forget plot]
table[row sep=crcr] {%
x	y\\
1.1875	0.00217805784732714\\
1.1875	0.00246572168384106\\
1.3125	0.00246572168384106\\
1.3125	0.00217805784732714\\
1.1875	0.00217805784732714\\
}--cycle;
\end{axis}
\end{tikzpicture}%

%% file: Electrodes_AtPotential.tex
%
%
\definecolor{mycolor1}{rgb}{0.0,0.7,0.2}%
\definecolor{mycolor2}{rgb}{0,0,1}%
\definecolor{mycolor3}{rgb}{1,1,1}%
\begin{tikzpicture}

\begin{axis}[%
width=3.05in,
height=1.9in,
at={(2.275in,1.015in)},
scale only axis,
xmin=0,
xmax=266,
xlabel style={font=\color{white!16!black}},
xlabel={Electrodes index},
ymin=-2e-05,
ymax=6e-05,
ylabel style={font=\color{white!16!black}},
ylabel={Potential},
axis background/.style={fill=white},
xmajorgrids,
ymajorgrids,
legend style={legend cell align=left, align=left, draw=white!16!black, fill=mycolor3}
]
\addplot [color=mycolor1, line width=1.5pt]
table[row sep=crcr]{%
1	2.49228714410515e-05\\
2	-2.66855795733756e-06\\
3	-1.96445920782138e-06\\
4	6.46488803221836e-06\\
5	1.16740665203392e-05\\
6	2.17787085472829e-05\\
7	3.39565245631891e-05\\
8	3.11380764507367e-05\\
9	1.61874173418051e-05\\
10	2.70057185332066e-07\\
11	-1.32385871466265e-05\\
12	-1.69797440135058e-05\\
13	1.44718896579156e-05\\
14	2.380811125398e-05\\
15	4.57859776491755e-05\\
16	5.09726334603963e-05\\
17	4.13056133683068e-05\\
18	1.44385019150375e-05\\
19	-4.32719015987352e-06\\
20	-1.50025336421834e-05\\
21	2.37755763453664e-05\\
22	4.01155673268321e-05\\
23	5.38015483085856e-05\\
24	5.38331972122601e-05\\
25	3.12642104144146e-05\\
26	1.02977657730841e-05\\
27	-9.95010727047706e-06\\
28	3.3480181523684e-05\\
29	4.10451926698601e-05\\
30	4.79714619536651e-05\\
31	3.72078693804806e-05\\
32	1.63143877831915e-05\\
33	1.23866565145866e-06\\
34	2.93639760731124e-05\\
35	3.10799819347239e-05\\
36	2.6862237006662e-05\\
37	1.74309731372524e-05\\
38	5.63466586451249e-06\\
39	1.85501291129389e-05\\
40	1.72078726693094e-05\\
41	1.20659262448153e-05\\
42	4.40708658413047e-06\\
43	-2.5872695582845e-06\\
44	-9.60956577295471e-06\\
45	-1.40612089599784e-05\\
46	-1.60937172338365e-05\\
47	-1.65399910481833e-05\\
48	8.82051443792857e-06\\
49	5.90818496203401e-06\\
50	2.02009985769184e-06\\
51	-2.9525968092462e-06\\
52	-8.93430895332974e-06\\
53	-1.2543789156685e-05\\
54	-1.49158794168587e-05\\
55	-1.56322930165879e-05\\
56	3.28800710286741e-06\\
57	7.5458423500028e-07\\
58	-3.57407398456248e-06\\
59	-7.38652274620683e-06\\
60	-1.09796697365406e-05\\
61	-1.32938567529256e-05\\
62	-1.40474824211365e-05\\
63	-8.86705278621572e-07\\
64	-3.3841091092721e-06\\
65	-6.62879064678641e-06\\
66	-8.9387200142162e-06\\
67	-1.13233408371498e-05\\
68	-1.27579405174032e-05\\
69	-1.13103506365913e-06\\
70	-3.17679578795944e-06\\
71	-5.55192837298256e-06\\
72	-7.92577493380328e-06\\
73	-9.78637610311953e-06\\
74	-1.11003824709661e-05\\
75	-1.52925243616162e-06\\
76	-6.67341980567878e-06\\
77	-9.15765250130816e-06\\
78	-1.03347067586808e-05\\
79	-1.17498881091722e-05\\
80	-1.31742840046312e-05\\
81	-1.41515828821869e-05\\
82	-1.53850949493822e-05\\
83	-1.62055165846396e-05\\
84	-1.80908799711033e-06\\
85	-6.22110420632159e-06\\
86	-8.05742046709594e-06\\
87	-9.51789513087788e-06\\
88	-1.07874140576532e-05\\
89	-1.21481497678264e-05\\
90	-1.35122829794855e-05\\
91	-1.44469616222823e-05\\
92	-1.52042291464019e-05\\
93	-1.59248527357925e-06\\
94	-2.62455749196073e-06\\
95	-4.51218677336602e-06\\
96	-5.84204108614207e-06\\
97	-7.02930812311244e-06\\
98	-8.71134181221032e-06\\
99	-1.00891321457692e-05\\
100	-1.14051031661752e-05\\
101	-1.2302502387374e-05\\
102	-1.33573843391134e-05\\
103	-1.31254613331704e-05\\
104	-2.17986703827531e-06\\
105	-3.85999506951211e-06\\
106	-5.36911112854861e-06\\
107	-6.6043831228703e-06\\
108	-7.94019884080864e-06\\
109	-8.97062720853687e-06\\
110	-9.96637559556497e-06\\
111	-1.15169789811911e-05\\
112	-1.24617526893368e-05\\
113	-3.10683267787472e-06\\
114	-4.45465276254116e-06\\
115	-5.8475587722336e-06\\
116	-7.06326189888718e-06\\
117	-8.12239960623163e-06\\
118	-9.23069654674428e-06\\
119	-1.00883175784663e-05\\
120	-1.10873663287197e-05\\
121	-1.21160630592587e-05\\
122	-3.61665471367897e-06\\
123	-5.17489302245794e-06\\
124	-6.49943075259444e-06\\
125	-7.49609330329189e-06\\
126	-8.22205107522747e-06\\
127	-9.21953034379477e-06\\
128	-1.02232680588595e-05\\
129	-1.10484093835149e-05\\
130	-1.19760410555069e-05\\
131	-1.31117262206477e-05\\
132	-1.40357426708494e-05\\
133	-1.50633058664394e-05\\
134	-1.61079594244732e-05\\
135	-4.45870800218923e-06\\
136	-5.4895821390603e-06\\
137	-6.90115025777958e-06\\
138	-7.36898020599682e-06\\
139	-7.97986952977604e-06\\
140	-9.11302769854256e-06\\
141	-9.80891261986719e-06\\
142	-1.07940030607044e-05\\
143	-1.17373235665317e-05\\
144	-1.28508168347582e-05\\
145	-1.36930710214999e-05\\
146	-1.47257980768748e-05\\
147	-4.47003642444449e-06\\
148	-5.24615801487826e-06\\
149	-6.48662527506487e-06\\
150	-7.44936051130518e-06\\
151	-7.95335538868242e-06\\
152	-8.93212817563e-06\\
153	-9.69325568305816e-06\\
154	-1.04596486438138e-05\\
155	-1.16496821464809e-05\\
156	-1.25684682386534e-05\\
157	-1.33179231030227e-05\\
158	-5.25237035021026e-06\\
159	-6.70058631991099e-06\\
160	-7.29581846056475e-06\\
161	-7.76679127359347e-06\\
162	-8.59470421791795e-06\\
163	-9.5324521916611e-06\\
164	-1.00217747948281e-05\\
165	-1.0827233634572e-05\\
166	-1.12333601381366e-05\\
167	-4.37090563847274e-06\\
168	-5.43248127787798e-06\\
169	-6.16547119181552e-06\\
170	-6.81720175022825e-06\\
171	-7.29944615227468e-06\\
172	-7.71139703173653e-06\\
173	-8.85459855102368e-06\\
174	-9.40285145567392e-06\\
175	-9.94053145740119e-06\\
176	-4.32805137540752e-06\\
177	-5.02768836292737e-06\\
178	-5.84872707319017e-06\\
179	-5.92808731589351e-06\\
180	-6.37028299258027e-06\\
181	-6.55952225071839e-06\\
182	-7.4333154632575e-06\\
183	-8.28716678493424e-06\\
184	-9.39601206777466e-06\\
185	-1.12992834228038e-05\\
186	-1.31903433354678e-05\\
187	-1.4741727225857e-05\\
188	-1.65956564576302e-05\\
189	-3.56827990051263e-06\\
190	-4.31615440367893e-06\\
191	-4.720768500059e-06\\
192	-4.96616043720358e-06\\
193	-5.31800786535457e-06\\
194	-5.63180234636095e-06\\
195	-6.35354665825593e-06\\
196	-6.73666838592744e-06\\
197	-8.46935452928851e-06\\
198	-9.42625320139738e-06\\
199	-1.04250412062171e-05\\
200	-3.08274687198997e-06\\
201	-3.5150254006132e-06\\
202	-3.66810976544241e-06\\
203	-4.29171192165798e-06\\
204	-4.76740006310784e-06\\
205	-3.68342875421486e-06\\
206	-4.5920657077201e-06\\
207	-3.34884607607964e-06\\
208	-3.49890407847486e-06\\
209	-2.32238917261722e-06\\
210	-2.6373704402717e-06\\
211	-1.82905911255997e-06\\
212	-1.82413085630235e-06\\
213	-4.59582109935897e-07\\
214	2.54420583432332e-06\\
215	4.48125641133948e-06\\
216	8.14110542331298e-06\\
217	-1.76582338000248e-06\\
218	-1.41247032609096e-06\\
219	-8.73388132824103e-07\\
220	-2.40099043479206e-07\\
221	7.13495183734193e-07\\
222	3.62982955313177e-06\\
223	7.33072305901512e-06\\
224	1.40749550531051e-05\\
225	1.59747447977338e-05\\
226	2.16936812427111e-06\\
227	3.88203625102492e-06\\
228	1.34952108343144e-06\\
229	2.90731851184266e-07\\
230	-4.46322975331933e-07\\
231	4.03666140471202e-06\\
232	2.69478618328177e-06\\
233	1.5660870412371e-06\\
234	7.71116519912146e-07\\
235	4.69516482676649e-06\\
236	3.3758751943989e-06\\
237	2.1168556080025e-06\\
238	1.25120541181085e-06\\
239	5.20539451938717e-06\\
240	3.23494945415291e-06\\
241	2.17843645294292e-06\\
242	4.58421157325147e-06\\
243	2.73470906264645e-06\\
244	1.87718445632092e-06\\
245	4.00027531739833e-06\\
246	2.41393533256315e-06\\
247	1.40625209705579e-06\\
248	8.81350397877223e-07\\
249	3.11785326851172e-06\\
250	1.19139969321021e-06\\
251	9.44565362445356e-07\\
252	2.61852254681747e-07\\
253	2.00718165905944e-06\\
254	9.16378222976833e-08\\
255	2.10633175660188e-07\\
256	-4.81574366260817e-07\\
};
\addplot [color=mycolor2, line width=1pt, draw=none, mark=x, mark options={solid, mycolor2}]
  table[row sep=crcr]{%
1	2.55502179434949e-05\\
2	-2.64690814270726e-06\\
3	-1.05388921929835e-06\\
4	7.26544668849717e-06\\
5	1.24199140046819e-05\\
6	2.25918213608936e-05\\
7	3.43398308818146e-05\\
8	3.17553556830532e-05\\
9	1.6762912201508e-05\\
10	8.00108890097082e-07\\
11	-1.29635173908045e-05\\
12	-1.67311150538439e-05\\
13	1.49235284283676e-05\\
14	2.40358240800066e-05\\
15	4.65904272327886e-05\\
16	5.19587377022923e-05\\
17	4.13356053185762e-05\\
18	1.49741661057048e-05\\
19	-4.24011293997263e-06\\
20	-1.42004422016276e-05\\
21	2.47647212550667e-05\\
22	4.01825135852299e-05\\
23	5.47409466704701e-05\\
24	5.38513747458968e-05\\
25	3.1948049028161e-05\\
26	1.10815022531673e-05\\
27	-9.41596970259433e-06\\
28	3.43655409746151e-05\\
29	4.19441975687663e-05\\
30	4.85973995797456e-05\\
31	3.73457383728931e-05\\
32	1.65321893769036e-05\\
33	1.42080672734909e-06\\
34	2.94057959370853e-05\\
35	3.11869235932742e-05\\
36	2.74786804917477e-05\\
37	1.83706341474134e-05\\
38	5.98912159547982e-06\\
39	1.89607582029984e-05\\
40	1.81922220862938e-05\\
41	1.30115054338506e-05\\
42	5.083731262564e-06\\
43	-1.59896729597122e-06\\
44	-8.84273438573811e-06\\
45	-1.37245096955869e-05\\
46	-1.5431335373437e-05\\
47	-1.6295825761393e-05\\
48	9.11602168876016e-06\\
49	6.58836333326452e-06\\
50	2.54794668811064e-06\\
51	-2.54100329583867e-06\\
52	-8.33167073529335e-06\\
53	-1.17932691007613e-05\\
54	-1.43323462425968e-05\\
55	-1.5080500501652e-05\\
56	3.87157772162512e-06\\
57	1.26640415495875e-06\\
58	-3.49148125756378e-06\\
59	-6.66695261134734e-06\\
60	-9.98351362524376e-06\\
61	-1.29393224479686e-05\\
62	-1.30762236059531e-05\\
63	-5.40256517321213e-07\\
64	-2.49756524751179e-06\\
65	-6.1740957817945e-06\\
66	-8.52529272519538e-06\\
67	-1.11056087687925e-05\\
68	-1.26322859300406e-05\\
69	-8.22120470092312e-07\\
70	-2.45069135629461e-06\\
71	-4.76905630000343e-06\\
72	-7.23198731881638e-06\\
73	-9.77657385085646e-06\\
74	-1.02571691329556e-05\\
75	-6.06920438365346e-07\\
76	-5.90246558500486e-06\\
77	-9.11499264537311e-06\\
78	-9.95652062163054e-06\\
79	-1.10455484846888e-05\\
80	-1.24447709591265e-05\\
81	-1.39273058115224e-05\\
82	-1.51160402176089e-05\\
83	-1.55324854196355e-05\\
84	-1.33159579938347e-06\\
85	-5.59738779365415e-06\\
86	-7.82097553445503e-06\\
87	-9.34077137637956e-06\\
88	-9.95777067195557e-06\\
89	-1.13812280998911e-05\\
90	-1.25778047063672e-05\\
91	-1.43390727171979e-05\\
92	-1.50220016408008e-05\\
93	-1.49338999115679e-06\\
94	-2.13479369003729e-06\\
95	-4.31894144289715e-06\\
96	-4.9461495123495e-06\\
97	-6.93021847343063e-06\\
98	-8.66717624044766e-06\\
99	-9.53183698996439e-06\\
100	-1.06326080989876e-05\\
101	-1.19905623298777e-05\\
102	-1.31784018597991e-05\\
103	-1.27865056549227e-05\\
104	-1.96972140123176e-06\\
105	-3.34984254974686e-06\\
106	-4.46274680528339e-06\\
107	-5.97545918421799e-06\\
108	-7.83866495199633e-06\\
109	-8.57977245581051e-06\\
110	-9.91175898034131e-06\\
111	-1.10156960679879e-05\\
112	-1.20300315174525e-05\\
113	-2.10927232836253e-06\\
114	-3.64305018154584e-06\\
115	-5.36190710233558e-06\\
116	-6.16881414331979e-06\\
117	-7.98485301146457e-06\\
118	-8.84069163229899e-06\\
119	-9.16096135346821e-06\\
120	-1.01698724963036e-05\\
121	-1.14024890476644e-05\\
122	-2.99831733005703e-06\\
123	-4.83160513221659e-06\\
124	-5.56340342590467e-06\\
125	-7.3713192626314e-06\\
126	-7.49146571372176e-06\\
127	-8.57305291136895e-06\\
128	-9.39011607319024e-06\\
129	-1.06501271552962e-05\\
130	-1.12262188461462e-05\\
131	-1.22765057101696e-05\\
132	-1.37132822734871e-05\\
133	-1.45110442495811e-05\\
134	-1.51288302920393e-05\\
135	-3.90939946917093e-06\\
136	-5.1591585294537e-06\\
137	-6.2816787026018e-06\\
138	-7.00834363499462e-06\\
139	-7.22335998627409e-06\\
140	-8.69912694985237e-06\\
141	-9.31656751548226e-06\\
142	-1.00992598275718e-05\\
143	-1.07645896814519e-05\\
144	-1.25230618742648e-05\\
145	-1.28552678384213e-05\\
146	-1.39867258496013e-05\\
147	-3.51586196806494e-06\\
148	-5.21423538537428e-06\\
149	-6.12975628888233e-06\\
150	-6.78670667701796e-06\\
151	-7.67185382953393e-06\\
152	-8.70174510831254e-06\\
153	-8.98212713187783e-06\\
154	-9.83507572682046e-06\\
155	-1.10590734935612e-05\\
156	-1.19080302723408e-05\\
157	-1.32703684299088e-05\\
158	-4.9035855417002e-06\\
159	-6.24924573955524e-06\\
160	-7.05491346344464e-06\\
161	-7.05174626029729e-06\\
162	-7.73852192591166e-06\\
163	-9.25094449654255e-06\\
164	-9.2907239651044e-06\\
165	-1.06894707420525e-05\\
166	-1.03966373563868e-05\\
167	-4.23230392273038e-06\\
168	-4.84427189248849e-06\\
169	-5.79931439136059e-06\\
170	-6.01044220556714e-06\\
171	-6.79566536649853e-06\\
172	-7.22180269301318e-06\\
173	-7.97754982763863e-06\\
174	-9.04970964273497e-06\\
175	-9.49108790082944e-06\\
176	-3.3645210885641e-06\\
177	-4.98539056501283e-06\\
178	-4.87576873904954e-06\\
179	-5.73888047277113e-06\\
180	-5.70316269254019e-06\\
181	-5.97308263603947e-06\\
182	-6.75820304685238e-06\\
183	-7.92614473573958e-06\\
184	-8.77573364070357e-06\\
185	-1.04881325377035e-05\\
186	-1.31710858580537e-05\\
187	-1.46578537175741e-05\\
188	-1.56208547904453e-05\\
189	-2.91693036809728e-06\\
190	-4.08491658751458e-06\\
191	-4.31727735693441e-06\\
192	-4.84413991895145e-06\\
193	-5.04956904395728e-06\\
194	-5.37395617624835e-06\\
195	-6.0218814195133e-06\\
196	-6.58443437306449e-06\\
197	-8.12134686957239e-06\\
198	-9.30459474708965e-06\\
199	-9.54088814846748e-06\\
200	-2.98846848192582e-06\\
201	-2.58498477450571e-06\\
202	-3.26908979640785e-06\\
203	-4.24431045962883e-06\\
204	-4.4250265599471e-06\\
205	-2.9474625953347e-06\\
206	-3.7973835503863e-06\\
207	-2.80394017783496e-06\\
208	-2.81268061549082e-06\\
209	-1.42875647668362e-06\\
210	-2.58257865035197e-06\\
211	-1.52539773284296e-06\\
212	-1.7779392999708e-06\\
213	-2.64105346302098e-07\\
214	3.26437163683985e-06\\
215	5.20300968483878e-06\\
216	9.01890449446451e-06\\
217	-1.18339041551507e-06\\
218	-1.34178599070431e-06\\
219	4.93564373649887e-08\\
220	5.60273048871901e-07\\
221	9.99442040154556e-07\\
222	4.17349278597314e-06\\
223	8.31549929598573e-06\\
224	1.47906331202317e-05\\
225	1.68137143950638e-05\\
226	2.60262868527906e-06\\
227	4.35266096663146e-06\\
228	1.91023449447599e-06\\
229	5.59823394715264e-07\\
230	3.02695492310013e-07\\
231	4.54054917817251e-06\\
232	3.34159584955455e-06\\
233	1.87383262326804e-06\\
234	9.0984115599029e-07\\
235	5.17073776056406e-06\\
236	3.73833447521765e-06\\
237	2.90496903640105e-06\\
238	2.03150123265496e-06\\
239	5.87390673364071e-06\\
240	3.36845331381422e-06\\
241	2.19999234014642e-06\\
242	5.14405227912398e-06\\
243	3.03552808071594e-06\\
244	2.81659417019438e-06\\
245	4.98117895344519e-06\\
246	2.70055572145741e-06\\
247	2.20707238400733e-06\\
248	1.77746174930983e-06\\
249	3.71537984532955e-06\\
250	2.07541642893403e-06\\
251	1.88829690364115e-06\\
252	8.1101034210165e-07\\
253	2.7355684836538e-06\\
254	6.68396120155693e-07\\
255	2.36490646743327e-07\\
256	-3.50433879760025e-08\\
};
\addlegendentry{FEM}
\addlegendentry{This work}
\end{axis}
\end{tikzpicture}%